\documentstyle[12pt]{article}
\newcommand{\be}{\begin{equation}}
\newcommand{\ee}{\end{equation}}
 \newcommand{\R}{{\rm I \hspace{-0.52ex} R}}
  \newcommand{\N}{{\rm I \hspace{-0.52ex} N}}

\newcommand{\bea}{\begin{eqnarray}}
\newcommand{\eea}{\end{eqnarray}}
\newcommand{\unop}{\rm 1 \hspace{-0.52ex} I}
\title{ QES systems, invariant spaces and
polynomials recursions.}
\author{Yves Brihaye,\\
{\it
 Universit\'e de Mons-Hainaut, Fac. Sciences
B-7000 Mons, Belgium},\\
Jean Ndimubandi, \\
{\it Universit\'e du Burundi, Fac. Sciences, B.P.2700,Bujumbura},\\
and\\
Bhabani Prasad Mandal,\\
{\it Department of Physics, Banaras Hindu University, Varanasi-221005,India.}}
\date{\today}
\begin{document}
\maketitle
\thispagestyle{empty}
\begin{abstract}
\par  
Let us denote ${\cal V}$, the
finite dimensional vector spaces of functions of the form 
$\psi(x) = p_n(x) + f(x) p_m(x)$
where $p_n(x)$ and $p_m(x)$ are arbitrary  polynomials of degree
at most $n$ and $m$ in the variable $x$ while $f(x)$ represents a
fixed  function of $x$.
 Conditions on $m,n$ and $f(x)$ are found  such that  families of
linear differential operators exist which preserve ${\cal V}$. A special 
emphasis is accorded to the cases where the set of differential operators
represents the envelopping algebra of some abstract algebra.
These operators can be transformed into linear matrix valued differential
operators. In the second part, such types of operators
are considered and a connection is established
between their solutions and series of polynomials-valued vectors obeying
three terms recurence relations.  When the operator is
quasi exactly solvable, it possesses a finite dimensional
invariant vector space. We study how this property leads to the
truncation of the polynomials series.
\end{abstract}
\section{Introduction}
Quasi Exactly Solvable (QES) operators are characterized by
linear differential operators
which preserve a finite-dimensional vector space ${\cal V}$  
of smooth functions \cite{tur1}.
In the case of operators
of one real variable the underlying vector space is often of the form 
${\cal V} = {\cal P}_n$ where
${\cal P}_n$  denotes the vector space of polynomials of degree
at most $n$ in the variable $x$.
In \cite{tur2} it is shown that the linear operators preserving
${\cal P}_n$ are generated by three
basic operators $j_-, j_0 , j_+$
(see Eq.(\ref{j}) below) which realize the algebra $sl(2,\R)$.
More general QES operators
can then be constructed by considering the elements of the enveloping
algebra of these generators, performing a change of variable 
and/or conjugating
the $j$'s with an invertible function, say
$g(x)$.  The effective invariant space is then the set of functions
of the form $g(x){\cal P}_n$.

In this paper we consider a more general situation.
 Let $m,n$ be two positive integers
  and let $f(x)$ be a sufficiently derivable function
 in a domain of the real line.
 Let ${\cal V}={\cal P}_n + f(x){\cal P}_m$
 be the vector space of functions of the form
 $p(x) + f(x) q(x)$  where $p(x)\in {\cal P}_n$,
 $q(x) \in {\cal P}_m$.

 We want to address the following questions~:
 What are the differential operators which preserve ${\cal V}$  ?
 and for which choice of $m,n,f(x)$ do these operators posses a relation
 with the enveloping algebra of some Lie (or "deformed`` Lie) algebra ?.
 This question generalizes the cases of monomials
 adressed in \cite{tur3} and more recently in \cite{dv}
At the moment  the question is, to our knowledge,
not solved in its generality
but we present a few non trivial solutions in the next section.


For all the cases we obtained, the final problem
of finding the algebraic modes of the scalar equations
leads to  2$\times$2-matrix QES operators preserving
the space of couples of polynomials with suitable degree.
On the other hand, a few example of systems of QES equations
are known and were studied in the past.

One problem that occur in computing the algebraic eigenvalues
is to diagonalize the Hamiltonian in a base of the its invariant
finite dimensional subspace. In the case of scalar QES
equations, this problem can be simplified by a technique
presented in \cite{bd}.
A few years ago, Bender and Dunne \cite{bd} have pointed out that
for a class of Schroedinger eigenvalue problems, the formal solution
constructed for generic values of the spectral parameter $E$
is the generating function of a set of orthogonal polynomials
$\{P_n(E) \}$.
This property is related to the fact
that the equation leads to a three term recursion relation on these
polynomials.
If the coupling constants of the Schroedinger operators
are choosen in such a way
that it becomes a quasi exactly solvable operator
\cite{tur,ush}, then
the polynomials obey a remarkable property, namely for $n$
larger than a fixed integer $N$ the polynomials $P_n(E)$ factorize
in the form $P_n(E) = P_N(E) \tilde P_{n-N}(E)$ where $P_N$ is
a common factor. The corresponding algebraic eigenvalues
associated with the "quasi-exact" property are the roots of $P_N(E)$.
The results of \cite{bd} was generalized and considered
in different frameworks in a series of papers \cite{fglr,khare1,khare2,khare3,khare4,ushveridze}.

However, to our knowledge, the approach of \cite{bd} has
not been applied to systems of quasi exactly solvable equations.
Apart from their mathematical interest, such systems appear
in a few context, e.g. in the construction of the doubly periodic
solutions of the Lam\'e equation \cite{arscott} (see e.g.
\cite{bg}) and in the stability analysis
of classical solutions available is some low dimensional field
theories \cite{bgkk,bakas,bt},.

It is the purpose of this letter to adapt the ideas of Bender and
Dunne to systems of coupled Schrodinger equations and to point
out that families of polynomials also appear in this context.
The eventuality of the orthogonality of these polynomials
is still an open question.

The paper is organized as follows. In Sect. 2 we review two
examples of well know coupled systems of QES equations, then
we discuss the cases of operators preversing  vector spaces
of the form $p(x) + f(x) q(x)$. The constuction of solutions
of the eigenvalue equation $H \psi = E \psi$ in a form of
polynomials series is adapted in Sect. 3 for the case of coupled
channels. It is shown that, along with \cite{bd} the $E$-dependant
coefficients of the series  obey three terms recurence relations.
Special emphasis is set on the relation between the finite dimensional
invariant space of $H$ and the truncation of the series.

\section{Systems of QES equations}
In this section, we review the most know 2$\times$2-matrix QES equations
and we study a class of scalar QES equations which  lead, 
after a suitable algebra, to
matrix operators acting on spaces of polynomials.

\subsection{Polynomial potential}
In this subsection, we study the Hamiltonian
\be
H(y) = -{d^2\over{dy^2}} \unop_2 + M_6(y)
\ee
where $M_6(y)$ is a $2\times 2$ hermitian matrix of the form
\begin{eqnarray}
M_6(y) &=& \lbrace 4p^2_2y^6+8p_1p_2y^4
+(4p^2_1-8mp_2+2(1-2\epsilon)p_2)y^2 \rbrace
\unop_2 \nonumber \\
&+& (8p_2 y^2+4p_1)\sigma_3-8mp_2\kappa_0\sigma_1
\label{m6}
\end{eqnarray}
where $\sigma_1, \sigma_3$ are the Pauli matrices,
$p_2,p_1,\kappa_0$ are free real parameters and $m$ is an integer.

It is known \cite{bh,zhdanov} that after the
standard ``gauge transformation'' of $H(y)$ with a factor
\be
\phi(y) = y^{\epsilon} \exp -\lbrace {p_2\over 2} y^4+p_1y^2\rbrace
\ee
and the change of variable $x=y^2$, 
the  new operator $\hat H(x)$
\be
     \hat H(x) = \phi^{-1}(x) H(y) \phi(x)\mid_{y=\sqrt x}
\ee
can be set in a form suitable for acting on polynomials.
However, the invariant space is revealed only after 
the supplementary transformation   \cite{bh}
\be
      \tilde H(x) = P^{-1} \hat H(x) P \ \ , \ \
      P = \unop_2 + \frac{\kappa_0}{2} \partial_x (\sigma_1 + \i \sigma_2)
\ee
is performed. After a calculation, we find
\be
       \tilde H(x) = - (4 x d_x^2 + 2 d_x)\unop
      + 8pm\kappa_0^2 \sigma_3 d_x  
\ee
\be
      + 8p {\rm diag}(J_+(m-2), J_+(m)) + 4\kappa_0(1+2mp \kappa^2) \sigma_+ d_x^2
      - 8m p \kappa_0 \sigma_-
\ee
which manifestly preserves the space of couples of polynomials
of the form \\
$(p_{m-2}(x),q_m(x))^t$. Here we choose $\epsilon = 0$
and $p_1=0$ for simplicity.
If the parameter $\epsilon$ is choosen as an arbitrary real number,
then the initial potential $M_6$ acquires
 a supplementary term of the form
$\epsilon (\epsilon-1)/y^2$.

 \subsection{Lam\'e type potential.}
\par As a second example, we consider the family of operators
\be
\label{lame}
H(z) = -{d^2\over{dz^2}} +
\left[\begin{array}{cc}
Ak^2{\rm{sn}}^2+\delta (1+k^2)/2 &2\theta k {\rm{cn}}\ {\rm{dn}}\\
2\theta k {\rm{cn}}\ {\rm{dn}} &Ck^2{\rm{sn}}^2-\delta (1+k^2)/2
\end{array}\right]
\ee
where $A,C,\delta, \theta$ are constants while ${\rm{sn}}, {\rm{cn}},{\rm{dn}}$ respectively abbreviate
the Jacobi elliptic functions of argument $z$
and modulus $k$ \cite{arscott}
\be
{\rm{sn}}(z,k) \quad , \quad {\rm{cn}} (z,k)\quad , \quad {\rm{dn}}(z,k)
\ \ .
\ee
These functions are periodic with period $4K(k),
4K(k), 2K(k)$ respectively ($K(k)$
is the complete elliptic integral of the first type). The above
hamiltonian is therefore to be considered on the Hilbert space of
periodic functions on $[0,4K(k)]$. 
For completeness, we mention the properties of the Jacobi
functions  which are needed in the calculations 
\be
{\rm{cn}}^2+{\rm{sn}}^2=1\quad , \quad {\rm{dn}}^2+k^2{\rm{sn}}^2=1
\ee
\be
{d\over{dz}}{\rm{sn}} = {\rm{cn}}\  {\rm{dn}}\quad ,
\quad {d\over{dz}} {\rm{cn}} = -{\rm{sn}}\  {\rm{dn}}\quad
, \quad {d\over{dz}} {\rm{dn}} = -k^2 {\rm{sn}}\  {\rm{cn}}
\ee

The relevant change of variable which 
eliminates the transcendental functions sn, cn, dn from (\ref{lame})
in favor of algebraic expressions is (for $k$ fixed)
\be
x={\rm{sn}}^2(z,k)
\ee
In particular the second derivative term in (\ref{lame})
becomes
\be
      {d^2 \over dz^2} =
4x(1-x)(1-k^2x) {d^2 \over d^2 x}
+ 2(3 k^2 x^2 - 2 (1+ k^2)x + 1){d \over d x}
\ee
Several possibilities of extracting prefactors then lead
to equivalent forms of (\ref{lame}), say $\tilde H(x)$,
which are matrix operators build with the derivative $d/dx$
and polynomial coefficients in $x$.
The requirement that $\tilde H(x)$ preserves a finite dimensional vector
space 
leads to two possible sets of
values for
$A,C,\theta$ (see \cite{bh} for details).
Here we will discuss the set determined by
\noindent $A=4m^2+6m+3-\delta$\\
$C=4m^2+6m+3+\delta$\\
$\theta = {1\over 2} [(4m+3)^2-\delta^2]^{1\over 2}$\\
The parameter $\delta$ remains free, and also $k$ which fixes the period
of the potential.
The results can be generalized easily to the other set \cite{bh}.

Corresponding to these values,
four invariant vector spaces are available. We will study only
one of them, refering, again, to \cite{bh} for the three others.

In order to present the change of variable we
conveniently define
\be
R = {4m+3-\delta\over{4m+3+\delta}},
\ee
We have then\\
\begin{eqnarray}
{\cal V}&=&
\left(\begin{array}{cc}
1 &0\\
0 &{\rm{cn}}\ {\rm{dn}}
\end{array}\right)
\left(\begin{array}{cc}
1 &\kappa x\\
0 &1
\end{array}\right)
\left(\begin{array}{c}
{\cal P}(m)\\
{\cal P}(m)
\end{array}\right)
\quad , \quad \kappa^2=k^2R
\end{eqnarray}\\

After performing the change of variable and the change of function,
the components of the operator (\ref{lame})
take the following form
\begin{eqnarray}
	\tilde H_{11}&=&-4x k^2(D+m+\frac{1}{2})(D-m)+(k^2+1)
        (4D^2+\frac{\delta}{2})-2(1+2D)d \ \ ,   \nonumber     \\
\tilde H_{12}&=&4x\kappa(k^2+1)(D-m)+\kappa(-8D+\delta+4m+1) \ \ ,
 \nonumber \\
\tilde H_{21}&=&\kappa(\delta+4m+3) \ \ ,  \nonumber \\
\tilde H_{22}&=&-4x k^2(D+m+\frac{5}{2})(D-m) \nonumber \\
&+&(k^2+1) (4D^2+2D+1-\frac{\delta}{2})-
2(1+2D)d \ \ ,
\end{eqnarray}
where $d \equiv d/dx$ and $D = xd$.
The operator $\tilde H$ obviously preserves $({\cal P}(m),({\cal P}(m))$.

\subsection{Scalar QES operators preserving $P_n \oplus f P_m$}
In this subsection, we investigate
\cite{brihaye} the linear operators preserving
vectors spaces of the form  $P_n \oplus f P_m$ and we construct 
several forms of the continuous function $f(x)$, together with
the associated values of the integers $m,n$ for which the 
operators of interest are the envelopping algebra
of some Lie (or deformed Lie) Algebra.

\subsubsection{Case $f(x) = 0$}
This is off course the well known case of \cite{tur1,tur2}. The
relevant operators
read
\be
\label{j}
     j_+(n) = x( x \frac{d}{dx} - n) \ \ , \ \
     j_0(n) = ( x \frac{d}{dx} - \frac{n}{2}) \ \ , \ \
     j_- = \frac{d}{dx}
\ee
and represent the three generators of $sl(2,\R)$. Most of known
one-dimensional QES equations are build with these operators.
For later convenience, we further define a family of equivalent
realizations of $sl(2,\R)$  by means of the conjugated operators
$k_{\epsilon}(a)\equiv x^a j_{\epsilon} x^{-a}$
for $\epsilon = +,0,-$ and $a$ is a real number.

\subsubsection{Case $f(x) = x^a$}
The general cases of vector spaces constructed over monomials
was first adressed
in \cite{tur3} and the particular subcase $f(x) = x^a$ was
reconsidered recently \cite{dv}.
The corresponding vector space  was
denoted $V^{(1)}$ in \cite{dv};
here we will reconsider this case and extend the discussion
of the operators
which leave it invariant. For later convenience,
it is usefull to introduce more precise notations, setting
${\cal P}_n \equiv {\cal P}(n,x) $ and
\begin{eqnarray}
\label{v1}
V^{(1)} \equiv  V^{(1)}(N,s,a,x) &=& {\cal P}(n,x)
+ x^a {\cal P}(m,x) \nonumber \\
&=& {\rm span} \{1,x,x^2,\dots ,x^n; x^a, x^{a+1}, \dots ,x^{a+m} \}
\nonumber \\
&=& V_1^{(1)} \oplus V_2^{(1)}
\end{eqnarray}
in passing, note that the notations of
\cite{dv} are  $n = s$ and $m = N-s-2$.

The vector space above is clearly constructed as the
direct sum of two subspaces.
As pointed out in \cite{tur3,dv}
three independent, second order differential operators
can be constructed which preserve
the vector space $V^{(1)}$. Writing these operators in the form
\begin{eqnarray}
      &J_+ &= x (x \frac{d}{dx} - n)
      (x \frac{d}{dx} - (m+a))  \nonumber \\
\label{jj}
      &J_0 &=  (x \frac{d}{dx} - \frac{m+n+1}{2})    \\
      &J_- &=  (x \frac{d}{dx} + 1 - a) \frac{d}{dx} \nonumber
\end{eqnarray}
makes it obvious that they preserve  $V^{(1)}$.

These operators close under the commutator
into a polynomial deformation of the $sl(2,\R)$ algebra:
\begin{eqnarray}
\label{nlalgebra}
 &[ J_0 , J_{\pm}] &= \pm J_{\pm} \nonumber \\
 &[ J_{+} , J_{-}] &= \alpha J_0 ^3 + \beta J_0 ^2 + \gamma J_0 + \delta
 \end{eqnarray}
where $\alpha, \beta, \gamma, \delta$ are constants  given in \cite{dv}.

Clearly the operators (\ref{jj}) leave
separately invariant  two vector spaces
$V_1^{(1)}$ and  $V_1^{(1)}$ entering in (\ref{v1}).
In the language of representations  they act reducibly on $V^{(1)}$.
However,  operators can be constructed which
preserve $V^{(1)}$ while
mixing the two subspaces. The form of these supplementary operators
is different according
to the fact that the number $a$ is an integer or not;
we now adress these two cases
separately.

\underline{Case $a\in \R$  }

In order to construct the operators which mix
$V_1^{(1)}$ and $V_2^{(1)}$, we first define
\begin{eqnarray}
&K &= (D-n)(D-n+1) \dots D  \ \ , \ \ D \equiv x\frac{d}{dx} \nonumber \\
&K' &= (D-m-a)(D-m-a+1) \dots (D-a)
\end{eqnarray}
which  belong to the kernals of the subvector spaces
${\cal P}_n$ and  $x^a {\cal P}_m$ of  $V^{(1)}$ respectively.
Notice that the products $j_\epsilon \tilde K$ and
$k_{\epsilon}(a) K$ (with $\epsilon = 0, \pm$)
also preserve the vector space.  For generic values
of $m,n$ these operators contain more than second derivatives
and, as so, they were not considered in \cite{dv}.

In order to construct the operators which
mix the two vector subspaces entering in ${\cal V}$,
we first have to construct the operators which transform a generic
element of ${\cal P}_m$ into an element of ${\cal P}_n$
and vice-versa.
 In \cite{bk} it is shown that these operators are of the form
\begin{eqnarray}
&q_{\alpha} &= x^{\alpha} \ \ \ , \ \ \alpha = 0,1,\dots, \Delta \ \\
&\overline q_{\alpha}&= \prod_{j=0}^{\alpha-1}(D-(p+1-\Delta)-j)
(\frac{d}{dx})^{\Delta-\alpha}
\end{eqnarray}
 where $\Delta \equiv \vert m-n \vert$ , $p \equiv $max$\{m,n\}$

The operators preserving ${\cal V}$ while exchanging the two
subspaces can finally be constructed by means of
\be
Q_{\alpha} = q_{\alpha} x^{-a} K \ \ , \ \
\overline Q_{\alpha} =  x^a \overline q_{\alpha}  K' \ \  , \ \
\alpha = 0,1,\dots, \Delta.
\ee
Here we assumed $n \leq m$, the case $n \geq m$ is
obtained by exchanging $q_{\alpha}$ with $ \overline q_{\alpha}$
in the formula above.

 It can be checked easily that
 $Q_{\alpha}$, transform a vector of the form $p_n + x^a q_m$ into
 a vector of the form $\tilde q_n \in {\cal P}_n$
 while  $\overline Q_{\alpha}$
 transforms the same vector
 into a vector of the form  $x^a \tilde p_m \in x^a {\cal P}_m$.

The generators constructed above are in one to one
correspondance with the 2$\times$2 matrix generators
preserving the direct sum of vector spaces
${\cal P}_m \oplus {\cal P}_n$ classified in \cite{bk}
although  their form is quite different (the same
notation is nevertheless used).
The commutation relations
(defining a normal order) which the generators fullfill is also
drastically different as we shall
discuss now.
First of all it can be easily checked that all products of
operators $Q$ (and separately of $\overline Q$)
belong to the kernal of the full space $V^{(1)}$, so we can write
\be
     Q_{\alpha} Q_{\beta} = \overline Q_{\alpha} \overline Q_{\beta} = 0
\ee
which suggests that the operators $Q$'s and the $\overline Q$'s
play the role of fermionic generators,
in contrast to the $J$'s which  are bosonic
(note that the same distinction holds in the case \cite{bk}).

From now on, we assume $n=m$ in this section
(the evaluation of the commutators for generic values
of $m,n$ is straightforward but leads to even more involved
expressions)
and suppress the superflous index $\alpha$ on the
the fermionic operators. The commutation relations between
fermionic and bosonic generators leads to
\begin{eqnarray}
   &[Q,J_-] = (2a-n-1) j_- Q \  , \
   &[Q,J_+]=(2a+n+1) j_+ Q    \\
   &[\overline Q,J_-] = -(2a+n+1) k_-(a) \overline Q \  ,  \
   &[\overline Q,J_+]= -(2a-n-1) k_+(a) \overline Q  \nonumber
\end{eqnarray}
where the $j_{\pm}$ and $k_{\pm}(a)$ are defined in (\ref{j}).
These relations define a normal order but we notice that the right
hand side are not linear expressions of the generators choosen
as basic elements.
We also have
\be
        [Q, D ]  = (D+a) Q  \ \   , \ \
        [\overline Q, D]  = (D-a) \overline Q
\ee
This is to be contrasted with the problem studied in \cite{bk}
where, for the case $\Delta = 0$,
the $Q$ (and the $\overline Q$)
commute with the three bosonic generators,
forming finally an sl(2)$\times$ sl(2) algebra.
Here we see that the bosonic operators $J$ and
fermionic operators $Q, \overline Q$
do not close linearly under the commutator.
The commutators involve in fact extra factors which can
be expressed in terms of the
operators $j$ or $k(a)$ acting on the appropriate subspace
${\cal P}_m$. This defines a normal order among
the basic generators but makes the underlying algebraic
stucture (if any) non linear.  For completeness, we also
mention that the anti-commutator
$\{Q, \overline Q \}$ is a polynomial in $J_0$.

 \underline{Case $a \in \N$}

 Let us consider the case $a \equiv k \in \N_0$, with
 $n \leq k $ and assume for definiteness
 $ m-k \geq n$.
 Operators that preserve $V^{(1)}$ while exchanging
 some monomials of the subspace $V_1^{(1)}$  with some of
 $V_2^{(1)}$  (and vice versa) can be expressed as follows~:
 \be
    W_+ = x^k \prod_{j=0}^{k-1} ( D - k - m + j)
 \ee
 \be
    W_- = \frac {1}{x^k} \prod_{j=0}^{n} ( D - j)
          \prod_{i=1}^{k-n-1} (D - k - n - i)
 \ee
 These operators are both of order $k$, $W_+$ is of degree $k$
 while $W_-$ is of degree $-k$.
 When acting  on the monomial of Eq.(\ref{v1}), $W_+$
  transforms the $n+1$ monomials of $V_1^{(1)}$
 into the first $n+1$ monomials of $V_2^{(1)}$ and annihilates the
 $k$ monomials of highest degrees in $V_2^{(1)}$.  To the contrary
 $W_-$ annihilates the $n+1$ monomials of $V_1^{(1)}$ and shifts
 the $n+1$ monomials of lowest degrees of   $V_2^{(1)}$  into
 $V_1^{(1)}$.  Operators of the same type performing higher jumps
 can be constructed in a straighforward way;
 they are characterized by a higher order and higher
 degrees but we will not present them
 here.

 The two particular cases $k=n+1$ and $k=2$ can be further commented.
 In the case $k=n+1$ the space $V^{(1)}$ is just ${\cal P}_{m+n+1}$
 and the operators $W_+$, $W_-$ can be rewritten as
 \be
        W_+ = (j_+(m+n+1))^{n+1}   \ \ , \ \ W_- = (j_-)^{n+1}
 \ee
where $j_{\pm}$ are defined in (\ref{j}).
 Setting $k=2$ (and $n=0$ otherwise we fall on the case just mentionned),
 we see that the operators $W_{\pm}$ become
 second order  and coincide with the operators noted
 $T_2^{(+2)}$ ,  $T_2^{(-2)}$ in Sect. 4
 of the recent preprint \cite{guk}. With our notation they read
 \be
       W_+ = x^2(D-(m+2))(D-(m+1)) \ \ , \ \
       W_- = x^{-2} D (D-3)
 \ee

 A natural question which come out is to study whether  the
  non-linear algebra (\ref{nlalgebra}) is extended
  in a nice way by the supplementary
  operators $W_{\pm}$ and their higher order
  counterparts. So far, we have not found any interesting
  extended structure.
  For example, for $k=2,n=0$ , we computed~:
  \begin{eqnarray}
  &[W_+ , J_+] &= -2 x^3 (D-(m+2))(D-(m+1))(D-m) \nonumber \\
  &[W_+ , J_-] &= -6 x D (D-(m+2)) (D-\frac{2}{3}(m+2))
   \end{eqnarray}
 which just show that the commutators close within the envelopping
 algebra of the $V^{(1)}$ preserving operators
 but would need more investigation to be confirmed as an abstract
 algebraic stucture.

We end up this section by  mentionning that
the two other vector spaces constructed in \cite{dv} and
denoted $V^{(a-1)}$ and $V^{(a)}$
can in fact be related to $V^{(1)}$
by means of the following relations~:
\be
V^{(a-1)}(x)   = V^{(1)} (N, s=0 , \frac{1}{a-1} , x^{a-1})
\ee
\be
V^{(a)}(x)   = V^{(1)}   (N, s, \frac{1}{a} , x^{a} )
\ee
Off course the operators  preserving them
can be  obtained
from the operators above (\ref{jj}) after
a suitable change of variable and the results above
can easily be extended to  these vector spaces.
\subsubsection{Case $f(x) = \sqrt{p_2(x)}$, $m=n-1$}
Here, $p_2(x)$ denotes a polynomial of degree 2 in $x$, we
take it in the canonical form  $p_2(x) = (1-x)(1-\lambda x)$.
In the case $m = n-1$, three basic operators can be contructed which
preserve ${\cal V}$ in the case $\lambda=-1$; they are of 
the form
\begin{eqnarray}
\label{sop}
       S_1 &=& n x + p_2 \frac{d}{dx}    \nonumber \\
       S_2 &=& \sqrt{p_2}(n x - x \frac{d}{dx})     \\
       S_3 &=& \sqrt{p_2}(\frac{d}{dx})  \nonumber
\end{eqnarray}
and obey the commutation relations of so(3). The operator corresponding to$\lambda \neq -1$ can be constructed from the three given above by a suitable affine transformation of the variable x.The family
of operators preserving ${\cal V}$ is in this case the enveloping
algebra of the Lie algebra of SO(3) in the realization above.
Two particular cases are worth to be pointed out~:
\begin{itemize}
 \item{ $\lambda =-1$}  \\
Using the variable $x = \cos(\phi)$, the vector space ${\cal V}$
can be re-expressed is the form
 \be
    {\cal V} = {\rm span} \{
    \cos(n \phi), \sin(n \phi), \cos((n-1)\phi),
    \sin((n-1)\phi), \dots
     \}
 \ee
 and the operators $S_a$ above can be expressed 
 in terms of trigonometric functions.
 Exemples of QES equations of this type were studied
 in \cite{uz} in relation with spin systems.
 \item{ $\lambda = k^2$ }
 Using the variable $x = {\rm sn}(z,k)$, (with   ${\rm sn}(z,k)$
 denoting the Jacobi elliptic function of modulus $k$, 
 $0 \leq k^2  \leq 1$), and
 considering the Lam\'e equation~:
 \be
 - \frac{d^2 \psi}{dz^2} + N(N+1)k^2 {\rm sn}^2(z,k)\psi = E \psi
 \ee
 It in known (see e.g. \cite{arscott})  that doubly periodic solutions
 exist if $N$ is a semi integer. 
 If $ N = (2n+1)/2$ these solutions are of the form
 \be
 \label{lame2}
    \psi(z) = \sqrt{{\rm cn}(z,k) + {\rm dn}(z,k)}
    (p_n(x) + {\rm cn}(z,k){\rm dn}(z,k) p_{n-1}(x))
 \ee
where ${\rm cn}(z,k), {\rm dn(z,k)}$ denote the other Jacobi
elliptic functions.
The second factor of this expression is exactly an element
of the vector space under consideration.
The relations between the
doubly periodic solutions  of the Lam\'e equation
and QES operators was pointed out in \cite{bg}.
\end{itemize}
\subsubsection{Case $f(x) =\sqrt{(1-x)/(1-\lambda x)}$}
For this form of $f(x)$, the vector space ${\cal V}$
is preserved by the operators $\tilde{S_a}$ with\\
$$\tilde{S_a}=\frac{1}{\sqrt{1-\lambda x}}S_a\sqrt{1-\lambda x}$$\\
provided $m=n$.

Two cases are worth considering, in complete paralelism
with Sect. 2.3.3:
\begin{itemize}
\item{$\lambda = -1$.}
Using the new variable $x =\cos \phi$, and using the identity
$\tan(\phi/2) = \sqrt{(1-x)/(1+x)}$ the vector space
${\cal V}$ can be reexpressed is the form
\be
{\cal V} = {\rm span}\{
\cos(\frac{2n+1}{2} \phi),
\sin(\frac{2n+1}{2} \phi),
\cos(\frac{2n-1}{2} \phi),
\sin(\frac{2n-1}{2} \phi), \dots
\}
\ee
and exemples of QES operators having
solutions in this vector space are presented in \cite{uz}.
\item{$\lambda = k^2$.}
 Again, in this case, the variable  $x = {\rm sn}(z,k)$
 is usefull and the doubly periodic solutions 
 of the Lam\'e equation (\ref{lame})
 corresponding to $N=(2n+3)/2$ of the form \cite{bg}
 \be
    \psi(z) = \sqrt{{\rm cn}(z,k) + {\rm dn}(z,k)}
    ( {\rm cn}(z,k) p_n(x) + {\rm dn}(z,k) q_{n}(x))
 \ee
 provide  examples of QES solutions constructed in the space under
 consideration.
\end{itemize}
\subsection{Relations with matrix operators}
The scalar operators constructed in the previous subsection
are equivalent to matrix valued differential operators acting
on polynomial-valued vector. This observation is rather trivial
but we illustrate the statement by mean of the case of Sect. 2.3.3.

Let us now reconsider the spaces of the form ${\cal V} =p_n+ f(x) q_m$
invariant through the symmetries $S_1,S_2$ and $S_3$ introduced above.
A closed link can be obviously made between
the later operators with matricials one which can be expressed in
terms of the usual generators $J_{\pm}(n)$ and $J_0$ of the $sl(2,\R)$
Lie algebra.\\ We obtain for$
f(x)=\sqrt{(1-x)(1-\lambda x)}$,$m=n-1$\\
$$\tilde S_1=
\left(\matrix{
J_--J_+(n)&0\cr
0&J_--J_+(n-1)\cr}\right)$$\\
$$ \tilde S_2=\left(\matrix{
0&-J_0(n/2)-x^2 J_+(n-1)\cr
-J_0(n/2)&0\cr}\right)$$\\
$$ \tilde S_3=\left(\matrix{
0&J_-- x(D+1)\cr
J_-&0\cr}\right)$$\\
\section{Recurence relations}
In this section, we will adapt the formulation of the QES
solution in terms of recurence relations \cite{bd} to the
case of matrix operator.
Let us assume that a $2*2$ matrix hamiltonian $H$
preserves a finite dimensional vector space.
Looking for solutions  of the form\\
\be
\label{series}
	\psi(y)=\psi_0(y)\sum_{n=0}^{\infty}P_n(E)y^n \ \ ,  \ \
P_n(E) \equiv \left(\begin{array}{c}
               p_{n}(E) \\
               q_n (E)
              \end{array} \right)
\ee\\
We will see that the eigenvalue equation $H \psi = E \psi$
leads to the a system of three terms recursion relations
        for the vectors $P_n(E)$ of the form:\\
	$$ P_{n+1} = (E \unop  + A) P_{n}+B P_{n-1}$$\\
where A and B are matrices depending on $n$ but independant
of the energy E. Moreover,  $B$ is diagonal and posseses
a zero eigenvalue for some specific value of $n$.
We will see how this property, together with the form of 
the recurence relations lead to a truncation of the formal series
(\ref{series}).

\subsection{Polynomial potential}
The solutions of the equation $$\tilde H(x) \Psi = E \psi$$
can be expressed in terms of formal series in the variable $x = y^2$
of the form\\
$\psi(x)=\exp(-p_2x^2/4)\sum_{n=0}^{\infty}P_n(E)x^n\\$
We can see after an algebra that the corresponding recurence relation has the form

\begin{equation}
  C_n \left(\begin{array}{c}
               p_{n+1} \\
               q_{n+2}
              \end{array} \right)
  = E \left(\begin{array}{c}
               p_n \\
               q_{n+1}
              \end{array} \right)
   + B_n \left(\begin{array}{c}
               p_{n-1} \\
               q_n
              \end{array} \right)            
\end{equation}

where\\
\be
\label{chbase}
B_n =8p
\left(\begin{array}{cc}
n-m-1 &0\\
0 &n-m
\end{array}\right)\\
  \ee
The form of $C_n$ can be obtained in a straightforward way but 
is irrelevant for our calculation.This equation can be 
solved recursively and it turns out that $q_0$ and $q_1$ 
are arbitrary parameters.
From the structure of matrix $B_n$,it turns out that 
$p_m$,$p_{m+1}$ and $q_{m+2}$ can be expressed
as linear combination of $p_{m-1}$ and $q_{m+1}$.
Therefore,considering the equations\\
$$p_{m-1}=0,q_{m+1}=0$$
as a linear system in the free parameters 
$q_0$,$q_1$, we obtain a polynomial condition on the
energy say $P(E)$. Those energies that fullfill $P(E)=0$
leads to a truncated series and therefore correspond to the 
quasi-exactly solvable solutions of the initial system.

\subsection{Lame type of potential}

After some calculus,we obtain a solution of the form

\be
 \psi(x) = \sum_{n=0}^{\infty} \frac{(-1)^n}{\Gamma(2n+1)}\textbf{P}_n(E)x^n\\
\ee 
Here,the matrices \textbf{A} and \textbf{B} are given by\\
$$
	\textbf{A}=
	\left(\begin{array}{cc}
	-(k^2+1)(4n^2+\frac{\delta}{2})&-\kappa(-8n+\delta+4m+1)\\
	-\kappa(\delta+4m+3)&-(k^2+1)(4n^2+2n+1-\frac{\delta}{2})\\
	\end{array}\right)$$

$$\textbf{B}=\left(\matrix{4k^2(n-m-1)(n+m-\frac{1}{2})&0\cr
	0&4k^2(n-m-1)(n+m+\frac{3}{2})\cr}\right)$$
	
The "initial" conditions write
$$P_1=\left(\matrix{
E-\frac{\delta}{2}(1+k^2) & -\kappa(\delta+4m+1)\cr
-\kappa(\delta+4m+3)& E-4(1+k^2)(1-\frac{\delta}{2})\cr}
\right)
P_0
$$
the vector $P_0$ being arbitrary.

We immediately observe that for $n=m+1$, the Matrix
$B$ vanish identically (its two eigenvalues are zero).
Therefore all the polynomials of the form $P_{m+j}$(j=2,3,...)
depend linearly of $P_{m+1}$. As a consequence, solving
the system of two equations $P_{m+1}=0$ as function of the
two arbitrary constants entering in $P_0$ leads to a
truncation of the series and a corresponding set of algebraic
eigenvalues.
This finally  leads to a factorization of all the polynomials
$P_{m+j}$(j=3,4,...) in terms of $P_{j+1}$.\\

\subsection{Bose-Hubbard Model}
In the above sections, we illustrated some new aspects of the
formulation of the QES property in terms of recursive polynomials.
Here we would like to present another exemple where still a new
feature of this formulation is revealed. It is connected to the
Bose-Hubbard (BH) model. It can be described by the
Schrodinger equation as
\begin{equation}
[-\frac{d^2}{dx^2} + V(x) ]\psi(x) = E_1\psi
\label{1}
\end{equation} where $V(x) =
\frac{1}{\gamma}\cosh^2(\sqrt{\gamma}x)
-(n+1)\cosh(\sqrt{\gamma}x)-\frac{1}{\gamma}-\gamma\frac{n}{2}(\frac{n}{2}-1)$
 and
$\frac{\gamma}{2}$ is the coefficient of the term $a^\dag a^\dag a
a $ term in the BH Hamiltonian of Ref. \cite{eilbeck,bdn}.

The Eq. (\ref{1}) can be written as
\begin{equation}
[-\frac{d^2}{dx^2} + ( \frac{1}{\alpha}\cosh \alpha x -M)^2]\psi
=E\psi \label{2}
\end{equation}
where, $\alpha =\sqrt{\gamma}, \ E=E_1+E_0,\ E_0
=\frac{(n+1)^2}{4}\alpha^2 +\frac{1}{\alpha^2}
+\frac{\alpha}{4}n(n-2),  \  M= \frac{(n+1)}{2}\alpha
$

Now to see the orthogonal polynomial associated with this model we
substitute,
\begin{equation}
\psi(x)= \exp{(-\frac{1}{\alpha^2}\cosh\alpha x)}\phi (x) \label{3}
\end{equation}
in the eq. (\ref{2}) to obtain,
\begin{equation}
\phi^{\prime\prime} -\frac{2}{\alpha}\sinh\alpha x \ \phi^{\prime}
+ [E-\frac{1}{\alpha^2}-M^2
+(\frac{2M}{\alpha}-\frac{1}{\alpha^2})\cosh\alpha x]\phi =0
\label{4}
\end{equation}
The Eq.(\ref{4}) further can be reduced to,
\begin{equation}
\alpha^2z(z+2)\phi^{\prime\prime}+[\alpha^2(z+1)-2z^2-4z]\
\phi^{\prime}+
[E-\frac{1}{\alpha^2}-M^2+(\frac{2M}{\alpha}-\frac{1}{\alpha^2})(z+1)]
\phi =0 \label{5}
\end{equation}
where $z=\cosh\alpha x-1$ and primes indicate derivatives with
respect to z. Now this equation has regular singular point at
$z=0$, therefore we seek a solution of the form,
\begin{equation}
\phi(z) =z^s f(z)
\end{equation}
By substituting this in Eq. (\ref{5}) and putting the coefficient s of
the term $z^{s-1}f(z)$ equal to zero, we obtain the indicial
equation for $s$ as $ 2s^2 = s$ . This implies $s$ can be either
$0$ or $\frac{1}{2}$. The differential equation in terms of f can
be written as,
\begin{eqnarray}
&&\alpha^2[(z+2)^2 -2(z+2)]\ f^{\prime\prime}
+[-(z+2)^2+(z+2)\{\alpha^2(2s+1)+2\}-\alpha^2]\ f^{\prime}
\nonumber \\ &+&[E-M^2+\alpha^2s^2-\frac{2M}{\alpha}+
(\frac{2M}{\alpha}-\frac{1}{\alpha^2})(z+2)]\ f=0 \label{6}
\end{eqnarray}

We further substitute
\begin{equation}
f= \sum_n \frac{R_n(E)}{n!}{(\frac{z+2}{2})}^{\frac{n}{2}}
\label{7}
\end{equation}
in the Eq. \ref{6} to obtain the three term recursion as
\begin{eqnarray}
\frac{\alpha^2}{4}R_{n+2}(E) &= & R_n(E)[ E +\frac{n^2\alpha^2}{4}
+sn\alpha^2+n +s^2 -M^2-\frac{2M}{\alpha}] \nonumber \\
&+&R_{n-2}(E) n(n-1)[\frac{2M}{\alpha}-\frac{1}{\alpha^2}-n-2]
\label{r1}
\end{eqnarray}
provided $ 2s^2 =s$. Thus  we have two sets of independent
solutions: the even states (i.e., states with even number of
nodes) for $s=0$ and the odd states for $s=\frac{1}{2}$. Note that
unlike Bender-Dunne ( or most other QES) cases, $s$ is not
contained in the potential and this is perhaps related to the fact
that for any integer value of $\tilde M$ with
$$
  \tilde M \equiv \frac{2 \alpha M - 1}{\alpha^2}
$$
the QES solutions corresponding
to both even and odd states are obtained. Also from Eq, \ref{r1}
we observe that the even and odd polynomials, $R_n(E)$ do not mix
with each other and hence we have two separate three-term
recursion relations depending on whether $n$ is odd or even. In
particular, it is easily shown that three term recursion relations
corresponding to the even and odd $n$ cases , respectively are
given by, $n\geq 1$

\begin{eqnarray}
\frac{\alpha^2}{4}P_n(E) &=& P_{n-1}(E) \ [ E+\alpha^2(n^2-2n+1
+2ns -2s)+2n-2+s^2-M^2-\frac{2M}{\alpha}]\nonumber \\ &+&
2P_{n-2}(E)\
(n-1)(2n-3)[\frac{2M}{\alpha}-\frac{1}{\alpha^2}-2n]\\
\frac{\alpha^2}{4}Q_n (E) &=& Q_{n-1}(E)\  [
E+\alpha^2(n^2-n+\frac{1}{4} +2ns
-2s)+2n-1+s^2-M^2-\frac{2M}{\alpha}]\nonumber \\ &+& 2Q_{n-2}(E) \
(n-1)(2n-1)[\frac{2M}{\alpha}-\frac{1}{\alpha^2}-2n-1]
\end{eqnarray}
with $P_0(E)=1\ Q_0(E) =1 $. These recursion relations generate a
set of monic polynomials and forms separately the complete set
orthogonal polynomials. These odd and even polynomials satisfied
the factorization properties of the Bender-Dunne polynomials. It
is easily seen that when $\tilde M$ is a positive integer, exact solutions
for first $\tilde M$ levels are obtained . In particular, if $\tilde{M}$ is odd
(even) integer, then solutions with even number of nodes ($s=0$)
are obtained when the coefficient of $P_{(n-2)}\ (Q_{(n-2)}$
vanishes. Similarly if $\tilde M$ is odd (even) integer, the solution with
odd number of nodes $(s=\frac{1}{2}$) are obtained when the
coefficient of $Q_{(n-2)}\ (P_{(n-2}) $  vanishes. Further for $\tilde M$
even (say $2k+2,\ k=0,1,2 \cdots $), half of levels, i.e. $k+1$
levels are obtained each from the zeros of the orthogonal
polynomials $P_{k+1}(E)$ and $Q_{k+1}(E)$ . On the other hand ,
when $\tilde{M}$ is odd  (say $2k+1,\ k= 0,1,2\cdots$ ) then $k+1$ and $k$
levels are obtained from the zeros of the orthogonal polynomials
$P_{k+1}(E) $ and $Q_k(E)$, respectively.

\section{Concluding remarks}
The examples of operators presented above give  evidences
of the difficulty to classify the coupled-channel
(or matrix) QES Schrodinger equations.
The way of constructing the QES  potential
$M_6$ in Sect. 2 further provides a clear link between the
approaches \cite{sz} and \cite{fgr} to this mathematical problem;
we hope that this note will  motivate further investigations
of it. We also presented a few new aspects of the recurence
relations for polynomials associated to QES operators. Especially
we formulated it for matrix operators. An open question is to
show that these vector-valued  polynomial are orthogonal with
respect to an appropriate measure.
We failed to find the matrix counterpart of the orthogonality theorem
\cite{chihara} for vector-valued polynomials.


\end{document}